\documentclass{emulateapj}
\usepackage{natbib}

\newcommand{\chandra}{{\it Chandra\/}}
\newcommand{\flux}{{erg~cm$^{-2}$~s$^{-1}$}}
\newcommand{\lum}{{erg~s$^{-1}$}}
\newcommand{\mlum}{{erg~s$^{-1}$~Hz$^{-1}$}}
\newcommand{\lnu}{{$\nu L_{\nu}$}}

\begin{document}

\title{Discovery of the Most-Distant Double-Peaked Emitter at
$z=1.369$}

\author{
B.~Luo,\altaffilmark{1}
W.~N.~Brandt,\altaffilmark{1}
J.~D.~Silverman,\altaffilmark{2}
I.~V.~Strateva,\altaffilmark{3}
F.~E.~Bauer,\altaffilmark{4}
P.~Capak,\altaffilmark{5}
J.~Kartaltepe,\altaffilmark{6}
B.~D.~Lehmer,\altaffilmark{7}
V.~Mainieri,\altaffilmark{8}
M.~Salvato,\altaffilmark{5}
G.~Szokoly,\altaffilmark{9,3}
D.~P.~Schneider,\altaffilmark{1}
\& C.~Vignali\altaffilmark{10}
}
\altaffiltext{1}{Department of Astronomy \& Astrophysics, 525 Davey Lab,
The Pennsylvania State University, University Park, PA 16802.}
\altaffiltext{2}{Institute of Astronomy, ETH Zurich, 8092, Zurich, Switzerland.}
\altaffiltext{3}{Max Planck Institute for Extraterrestrial Physics, Garching, 85741, Germany.}
\altaffiltext{4}{Columbia Astrophysics Laboratory, Columbia University,
Pupin Laboratories, 550 West 120th Street, New York, NY 10027.}
\altaffiltext{5}{California Institute of Technology, 1200 East California Boulevard, 
Pasadena, CA 91125.}
\altaffiltext{6}{Institute for Astronomy, University of Hawaii, 2680 Woodlawn 
Drive, Honolulu, HI 96822.}
\altaffiltext{7}{Department of Physics, Durham University,
Durham DH1 3LE, UK.}
\altaffiltext{8}{European Southern Observatory, Karl-Schwarzschild-Strasse 2,
Garching, D-85748, Germany.}
\altaffiltext{9}{E\"otv\"os University, Inst. of Physics, 
1117 Budapest, P\'azm\'any P. s. 1/A, Hungary.}
\altaffiltext{10}{Universit\'a di Bologna, Via Ranzani 1, Bologna, Italy.}

\begin{abstract}
We report the discovery of the most-distant
double-peaked emitter, \hbox{CXOECDFS} J033115.0$-$275518, at $z=1.369$. A Keck/DEIMOS
spectrum shows a clearly double-peaked broad \ion{Mg}{2} $\lambda2799$ emission line,
with ${\rm FWHM}\approx11\,000$~km~s$^{-1}$ for the line complex. 
The line profile can be well fit by
an elliptical relativistic Keplerian disk model. 
This is one of a handful of double-peaked emitters known to be a 
luminous quasar,
with excellent multiwavelength coverage and  
a high-quality X-ray spectrum.
CXOECDFS J033115.0$-$275518 is a 
radio-loud quasar with two radio lobes (FR II morphology) 
and a radio loudness of $f_{5~{\rm GHz}}/f_{\rm 4400~\AA }\approx429$. 
The X-ray spectrum can be
modeled by a power law with photon index 1.72
and no intrinsic absorption; the rest-frame \hbox{0.5--8.0~keV}
luminosity is $5.0\times10^{44}$~\lum. The spectral energy distribution (SED) 
of CXOECDFS J033115.0$-$275518 
has a shape typical for radio-loud quasars and
double-peaked emitters at lower redshift.
The local viscous energy released 
from the line-emitting region of the accretion disk
is probably insufficient to power the observed line flux, 
and external illumination of the disk appears to be
required. The presence of a big blue bump in the SED along with
the unexceptional X-ray spectrum suggest that 
the illumination cannot arise 
from a radiatively inefficient accretion flow.
\end{abstract}


\section{Introduction}

Double-peaked emitters are active galactic nuclei (AGNs) emitting
broad and double-peaked low-ionization lines. A survey of $\sim100$
radio-loud AGNs ($z<0.4$) suggests 
that $\sim20\%$ of these sources are double-peaked 
emitters \citep{Eracleous1994,Eracleous2003}. The frequency of  
double-peaked emitters
is much lower (\hbox{$\sim3\%$}) among the general population of $\sim3000$ 
low-redshift ($z<0.33$) AGNs
selected by the Sloan Digital Sky Survey \citep{Strateva2003}, 
and these double-peaked emitters are predominantly (76\%) radio-quiet. 
Since the discovery of the first examples of double-peaked emitters in the 
1980s
\citep{Oke1987,Chen1989a}, more than 150 such sources have been found.
Most have been identified based on their H$\alpha$, and
sometimes H$\beta$, lines. Spectroscopy of several of these
sources showed that 
the \ion{Mg}{2} $\lambda2799$ line\footnote{This \ion{Mg}{2} line is actually
a $\lambda2796/2803$ doublet. Due to line broadening, 
the two components are usually blended and cannot be resolved.
Thus we treat the doublet as a single line when discussing the 
observed emission feature throughout this paper.} 
also has a double-peaked profile similar 
to those
of H$\alpha$ and H$\beta$ \citep{Halpern1996,Eracleous2004}. Selection based on 
the \ion{Mg}{2} line profile could in principle find higher redshift candidates;
however, due to contamination from the 
underlying \ion{Fe}{2} and \ion{Fe}{3} emission-line complexes 
\citep[e.g.,][]{Wills1985} and possibly
\ion{Mg}{2} self-absorption, it is difficult to 
create a complete sample of double-peaked 
\ion{Mg}{2} $\lambda2799$ emitters, and only a few such objects have been reported
\citep{Strateva2003}. The highest redshift double-peaked emitters 
discovered to date
have $z\approx0.6$.

Observational results and basic physical considerations indicate that the 
most-plausible
origin of the double-peaked emission lines is the accretion disk  
\citep[e.g.,][]{Chen1989,Eracleous1995,Eracleous2003}. 
The line
profile can be well fit by emission from the outer regions of
a Keplerian disk, typically
at distances from the black hole of hundreds to thousands
of gravitational radii ($R_{\rm G}=GM/c^2$). In many cases, 
the viscous energy available
locally in the line-emitting region is insufficient to power the 
observed lines, and it has been suggested that these strong lines are 
produced by external illumination of the disk, probably from an
\hbox{X-ray-emitting} elevated structure in the center \citep[e.g.,][]{Chen1989a}.

In this paper, we report the discovery of the highest redshift double-peaked 
emitter known to date. This source, CXOECDFS J033115.0$-$275518
(hereafter J0331$-$2755), 
was detected as an \hbox{X-ray} 
point source in the $\approx 250$~ks 
observations of the Extended \chandra\ Deep Field-South (\hbox{E-CDF-S}; 
\citealt{Lehmer2005}), and was identified as a double-peaked \ion{Mg}{2}
emitter
at $z=1.369$ by Keck/DEIMOS optical spectroscopy. 
We adopt a cosmology with 
$H_0=70$~km~s$^{-1}$~Mpc$^{-1}$, $\Omega_{\rm M}=0.3$, 
and $\Omega_{\Lambda}=0.7$ throughout this paper.

\section{Keck Observations and Disk-Model Fit}

Optical spectroscopic observations of J0331$-$2755 were 
carried out using the
DEIMOS spectrograph \citep{Faber2003} on the 10 m Keck II telescope on January
15, 2007 (UT), as part of the E-CDF-S spectroscopic program (PIs: P.~Capak,
M. Salvato, J. Kartaltepe; Silverman et~al. 2009, in preparation).
We used the 600 l/mm grism and the GG455 filter. The wavelength coverage was
$\sim4650$--8580 \AA\ 
with a resolution of 3.5 \AA.
The seeing was 0\farcs6, and the airmass was 1.49. The total exposure time was
9000 s in five individual exposures. 
The wavelength-dependent response was corrected by an observation of a single
flux standard star while the overall normalization was set to match the 
COMBO-17 $R$-band magnitude \citep{Wolf2004,Wolf2008}.
The redshift of J0331$-$2755,
1.369, was determined using the \ion{Mg}{2} $\lambda2799$ and 
\ion{Ne}{5} $\lambda3426$ narrow lines (${\rm FWHM}\approx400$~km~s$^{-1}$).

\begin{figure}
\centerline{
\includegraphics[scale=0.5]{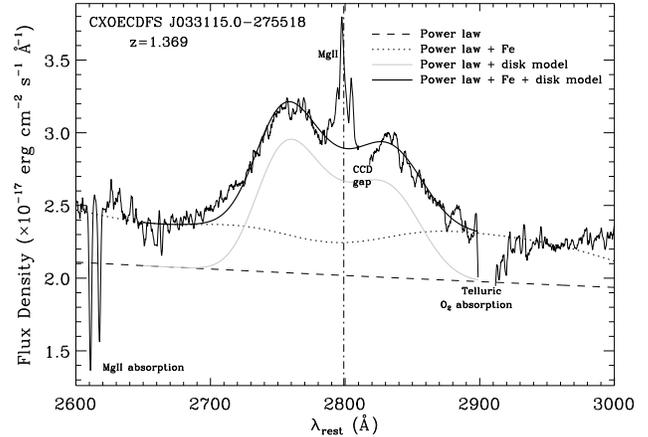}
}
\figcaption{Rest-frame NUV spectrum of J0331$-$2755 showing
the double-peaked \ion{Mg}{2}
line. 
The vertical dash-dotted line indicates the expected position of
\ion{Mg}{2} $\lambda2799$.
Two gaps in the spectrum are caused by
the DEIMOS CCDs and telluric ${\rm O_2}$ absorption.
Dashed, dotted, grey, and thick curves are different emission 
components used to fit 
the observed double-peaked \ion{Mg}{2} line profile, as indicated in the plot.
The rms noise of the spectrum 
is $1.3\times 10^{-18}$~erg~cm$^{-2}$~s$^{-1}$~\AA$^{-1}$.
\label{opspec}}
\end{figure}

The rest-frame near-UV (NUV) spectrum around the 
\ion{Mg}{2} $\lambda2799$ line is shown in Figure~\ref{opspec},
smoothed to a resolution of 3.6 \AA. The spectrum
shows a clearly double-peaked broad \ion{Mg}{2} line along with a central narrow
line, similar to previously discovered 
double-peaked H$\alpha$ or \ion{Mg}{2} emitters.
The \ion{Mg}{2} absorption doublet at $\sim2600$~\AA\ is likely produced 
by an intervening absorber at $z=1.21$, as inferred from the 
narrow velocity dispersion (${\rm FWHM}\approx300$~km~s$^{-1}$)
and large blueshift \citep[e.g.,][]{Ganguly2007}. 
An alternative interpretation of the doublet as arising in an AGN outflow
would require outflow velocities of $\sim20\,000$~km~s$^{-1}$.

Assuming that the J0331$-$2755 \ion{Mg}{2} line emission originates from the
accretion disk, we can use the line shape to determine a set of
parameters characterizing the emission region \citep[see][for a
description of line-profile accretion-disk
modeling]{Chen1989,Eracleous1995}. We start by subtracting the
underlying continuum and Fe emission-line complexes, as shown in
Figure~\ref{opspec}. The continuum is represented by a simple
power law, $F_\lambda\propto \lambda^{-1.6}$ \citep{Vandenberk2001}.
The Fe pseudo-continuum is modeled by the broadened Fe-emission
template of \citet{Vestergaard2001}, fit to the
$\sim2450$--2900\,\AA\ spectrum excluding the \ion{Mg}{2} line, where
we assumed that the Fe-line broadening is similar to that of the
\ion{Mg}{2} line complex, ${\rm FWHM}\approx11\,000$~km~s$^{-1}$. Such Fe-line
broadening could result if the line is emitted from the base of a wind
launched from the accretion disk. The double-peaked \ion{Mg}{2} line profile does
not differ much for any reasonable choice of the Fe-line
broadening (from 5000 to 15\,000~km~s$^{-1}$).

The continuum and Fe-line complex subtracted \ion{Mg}{2} profile of
J0331$-$2755 cannot be well fit by a circular relativistic Keplerian disk
model. In the absence of line-profile variability, which can help
distinguish between the different non-axisymmetric disk models, we
choose the elliptical disk model of \citet{Eracleous1995}, which has
the smallest number of extra free parameters (for a total of 7
disk-fit parameters). The model assumes an external source of
illumination represented by a power law with emissivity, $\epsilon
\propto R^{-q}$, and we fix $q=3$, equivalent to an illuminating
point source on the disk axis high above the disk. 
The central narrow-line part of the spectrum was excluded from the fit.
From the best-fit model, 
the disk inclination with respect to our line of sight is 
$i={22^{+8}_{-3}}$ degrees,
the inner and
outer radii of the emission ring are $R_1=200^{+200}_{-50}\,R_{\rm G}$ and
$R_2=900^{+600}_{-100}\,R_{\rm G}$,
the turbulent broadening parameter is
$\sigma=1900^{+400}_{-200}$~km~s$^{-1}$, and the disk ellipticity is 
$e=0.4^{+0.1}_{-0.2}$, with a
semi-major axis orientation of $\phi_0=110\pm20$ degrees with respect to our
line of sight. The integrated 
line flux is $F_{\rm Mg~II}=(9.3\pm0.2)\times10^{-16}$~\flux. 
These
emission-line region parameters are similar to those obtained for
lower-redshift double-peaked emitters, e.g.,
\citet{Eracleous2003}, \citet{Strateva2003}, and \citet{Strateva2008}.

\section{Multiwavelength Properties}

\begin{figure}
\centerline{
\includegraphics[scale=0.45]{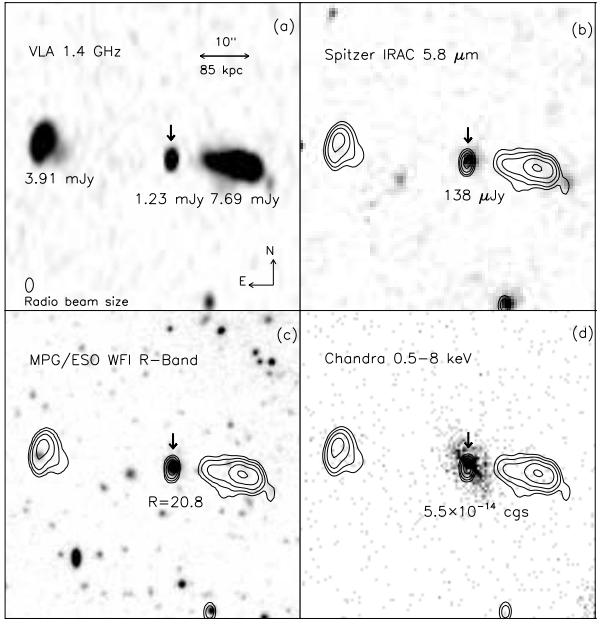}
}
\figcaption{
({\it a}) Radio 1.4 GHz, ({\it b}) IR 5.8 $\mu$m, ({\it c}) optical $R$-band, 
and ({\it d}) X-ray 0.5--8 keV images of J0331$-$2755. Each image
is 60\arcsec\ (0.51 Mpc at $z$=1.369) on a side. 
The last
three images are overlaid with radio contours, ranging from $\approx0.6$--90\% of the maximum pixel value following a logarithmic scale.
The downward arrows point to the position of J0331$-$2755.
The ellipse at the lower left corner of ({\it a}) shows the beam size of the radio
observations.
Flux density or flux in each band is indicated; the X-ray flux
is in units of \flux.
The apparent extension in the \chandra\ image is due to the large
point spread function at its location.
\label{img}}
\end{figure}

\begin{figure}
\centerline{
\includegraphics[scale=0.5]{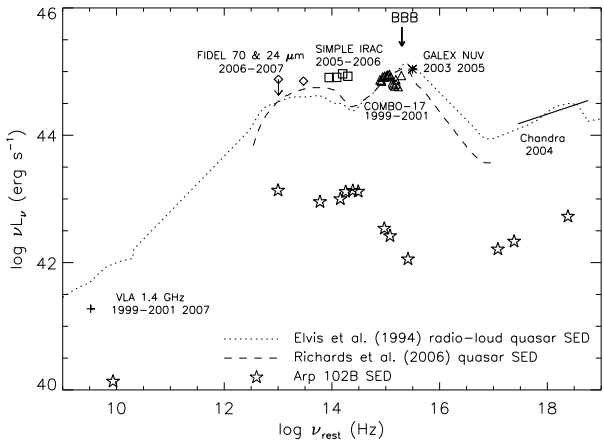}
}
\figcaption{Radio through X-ray SED of J0331$-$2755. 
The names of observatories/surveys and years of observations
are indicated for the data points. 
The dotted and dashed curves
show the \citet{Elvis1994} and \citet{Richards2006} 
SED templates, normalized to the COMBO-17
$R$-band data point, respectively. The broad-band SED of J0331$-$2755 is
in general agreement with typical radio-loud quasar SEDs. The Arp 102B
SED \citep[][and references therein]{Eracleous2003a,Strateva2008} 
is also shown for comparison. The J0331$-$2755 SED
and both of the SED templates show a big blue bump in the UV
(marked with a thick downward-pointing arrow), while 
the Arp 102B SED does not.
\label{sed}}
\end{figure}

The numerous multiwavelength deep surveys of the \hbox{E-CDF-S}
allow us to study the properties of J0331$-$2755 from
radio to \hbox{X-ray} wavelengths. 
Figure~\ref{img} presents the
radio, infrared (IR), optical, and X-ray images of J0331$-$2755.
A broad-band spectral energy distribution (SED) 
of the source is shown in Figure~\ref{sed}, with the majority of the SED 
data displayed in Table~\ref{phodata}.
It is one of the best-sampled double-peaked emitter SEDs, 
comparable to that of the prototype source Arp 102B \citep{Eracleous2003a,Strateva2008}.
Details of the 
broad-band properties are discussed below.


{\it Radio} --- J0331$-$2755 was observed by the
Very Large Array (VLA) at 1.4 GHz
in 1999--2000 \citep{Kellermann2008} and 
2007 \citep{Miller2008}.\footnote{http://taltos.pha.jhu.edu/$\sim$nmiller/vlaecdfs\_main.html.} The reported core
radio flux densities are $1.14\pm0.03$~mJy and $1.23\pm0.02$~mJy, respectively.
It was also detected by the Australia Telescope Compact Array (ATCA) at 1.4 GHz,
with a $\sim30\%$ higher flux density \citep{Rovilos2007}. 
Because of the low resolution (beam size 
$17\arcsec\times7\arcsec$), we do not use the ATCA results.
The average VLA flux density is plotted in Figure~\ref{sed}.
We show the radio image from \citet{Miller2008} in Figure~\ref{img}a. 
Most of the data were obtained when the VLA was in configuration A, and some
were obtained in configuration BnA. 
The beam size is $2.8\arcsec\times1.6\arcsec$ with a 
position angle near zero (i.e., N-S).
The source is associated with two brighter radio lobes, 
which extend to a few hundred
kpc away from the center, displaying an FR II morphology. 
The core and the lobes are sources 20, 19, and 21 in
the catalog of \citet{Miller2008}, and the total radio power 
at 1.4 GHz is $\sim1.5\times10^{33}$~\mlum.
According to the radio contours, 
there could also be a weak jet leading to the lobe
in the western feature.
We did not find any clear counterparts 
for the lobes/jet at other wavelengths, except for a faint optical source at 
the position of the eastern lobe. Moreover, these structures 
are more extended than
nearby point sources. Thus they are 
not likely physically associated with galaxies. The radio loudness parameter,
defined as $R=f_{5~{\rm GHz}}/f_{\rm 4400~\AA }$ \citep[the ratio of 
flux densities in the rest frame; e.g.,][]{Kellermann1989}, is computed 
using the 1.4 GHz and 914 nm COMBO-17 \citep{Wolf2004,Wolf2008} 
flux densities
with the assumption of a radio power-law slope of $\alpha_{\rm r}=-0.8$ 
and an optical power-law slope of $\alpha_{\rm o}=-0.4$ 
($F_\nu\propto \nu^{\alpha}$). The source
has a radio loudness of $R\approx429$, including the contributions from
the extended lobe emission.\footnote{The integrated radio flux density was 
used to compute the radio loudness in order
to be consistent with the typical definition in the literature. The 
radio loudness is $\sim41$ if only the radio flux from
the core component is taken into account.} 

{\it IR} --- The \hbox{E-CDF-S} was covered by the {\it Spitzer} Far Infrared Deep Extragalactic 
Legacy Survey (FIDEL) at 24 and 70~$\mu$m,\footnote{See
http://ssc.spitzer.caltech.edu/legacy/fidelhistory.html.}
and by the {\it Spitzer}
IRAC/MUSYC Public Legacy Survey in the \hbox{E-CDF-S} (SIMPLE)
at 3.6, 4.5, 5.8, and 
8.0 $\mu$m.\footnote{See http://ssc.spitzer.caltech.edu/legacy/simplehistory.html.}
J0331$-$2755 was detected at 3.6, 4.5, 5.8, 8.0, and 
24 $\mu$m with flux densities of 0.09, 0.12, 0.14, 0.19, and 0.50 mJy, respectively
(e.g., see Figure~\ref{img}b). 
It was not detected at 70 $\mu$m; we estimate the $2~\sigma$ flux-density
upper limit to be 1.5 mJy.

{\it Optical, UV} --- In the optical band, J0331$-$2755 is present in the 
COMBO-17 catalog \citep{Wolf2004,Wolf2008} and 
the deep MPG/ESO Wide Field Imager (WFI) 
$R$-band catalog \citep{Giavalisco2004}. It is outside the field-of-view of the 
Galaxy Evolution from Morphologies and SEDs (GEMS)
survey \citep{Caldwell2008}.
The $B$-band absolute AB magnitude is $M_{\rm B}=-23.4$.
The WFI $R$-band image is shown in
Figure~\ref{img}c and the 17-band photometry data points from the 
COMBO-17 observations are shown in Figure~\ref{sed}. The galactic extinction
in the optical band is small (a correction factor of $\sim1.02$ for the
$V$-band).
J0331$-$2755 was observed by {\it GALEX} in 2003 and 2005.\footnote{See 
http://galex.stsci.edu/GR4/.} The observed
NUV ($\lambda_{\rm eff}=2267$~\AA)
flux densities are $5.95\pm0.13$ and $7.69\pm0.23$ 
$\mu$Jy, brightening by $\sim30\%$ in two years. This variability amplitude is 
typical for double-peaked emitters \citep[e.g.,][]{Gezari2007,Strateva2008}.
The correction factor for the NUV Galactic extinction is $\sim1.07$,
following the Galactic extinction law of $A_{\rm NUV}=8.0E(B-V)$ 
\citep[e.g.,][]{Gildepaz2007}.
The source was not detected in the far-UV band 
($\lambda_{\rm eff}=1516$~\AA),
due to the Lyman break at rest-frame 912 \AA.
The average NUV flux density is shown in Figure~\ref{sed}.

{\it X-ray} --- J0331$-$2755 was detected during the \chandra\ \hbox{E-CDF-S} survey
in 2004, with an effective exposure time of $\approx210$ ks. It is located 
at the edge of the \hbox{E-CDF-S} field and has an off-axis angle of $\sim8\arcmin$.
The \chandra\ image is shown in Figure~\ref{img}d.
J0331$-$2755 has one of the best \hbox{X-ray} spectra available 
for double-peaked emitters; 
the number of \hbox{0.5--8.0~keV} source counts is 1048. 
We performed 
spectrum extraction on the reduced and cleaned level~2 event files
\citep{Lehmer2005} using the reduction tool
{\sc acis extract} (AE; \citealt{Broos2000}).
The spectrum was binned at a signal-to-noise ratio of 5 
using AE and then fit with XSPEC (Version 12.4.0;
\citealt{Arnaud1996}). The \hbox{0.5--8.0~keV} spectrum can be well modeled
with a power law modified by Galactic absorption, with $\chi^2=18.0$ 
for 22 degrees of freedom (see Fig. 4). We adopted a Galactic neutral 
hydrogen column density \hbox{$N_{\rm H,G}=8.8\times 10^{19}$~cm$^{-2}$} 
\citep{Stark1992}.
The best-fit photon
index is $\Gamma=1.72\pm0.10$; the uncertainties are quoted at 90\% 
confidence. This photon index is typical for radio-loud quasars, 
$\Gamma_{\rm RL}\approx1.6$--1.7 \citep[e.g.,][]{Reeves1997}.
The rest-frame unabsorbed \hbox{0.5--8.0~keV} 
luminosity is  
$5.0\times10^{44}$~\lum, well within the quasar regime. 
Intrinsic absorption ($N_{\rm H,i}$ at $z=1.369$) 
is not required to fit
the spectrum, with a 90\% confidence upper limit of
$1.09\times10^{22}$ cm$^{-2}$. 

\begin{figure}
\centerline{
\includegraphics[scale=0.5]{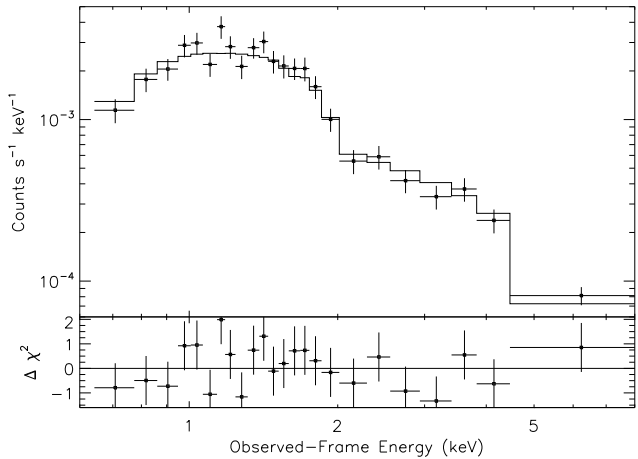}
}
\figcaption{X-ray spectrum of J0331$-$2755
overlaid with the best-fit model.
The bottom panel shows the deviation of the data from the model in units of
$\sigma$, with error bars of size unity. The spectrum can be modeled
with a simple power law modified by Galactic absorption; the best-fit photon
index for this model is $\Gamma=1.72\pm0.10$. 
\label{xspec}}
\end{figure}

To compare the SED of J0331$-$2755 to those for typical quasars, we show
in Figure~\ref{sed} the mean quasar SED from \citet{Richards2006} and the mean
radio-loud quasar SED from \citet{Elvis1994}, both normalized to the COMBO-17 
$R$-band flux of J0331$-$2755. The \citet{Richards2006} SED template 
is derived from a sample of radio-quiet sources, and
does not cover
the radio and \hbox{X-ray} bands. The optical--to--X-ray data agree reasonably well
with the mean quasar SEDs. The excess emission in the 
IR bands suggests a contribution from the host galaxy, similar to some other 
double-peaked emitters \citep[e.g.,][]{Strateva2008}. The 
weaker radio emission than the mean radio-loud SED probably arises because
we only include the radio emission from the core, while some of the 
\citet{Elvis1994} objects included extended radio emission when only
low-resolution observations were available.
As the multiwavelength data were not taken simultaneously,
variations in the fluxes at different wavebands are likely responsible for part of
the discrepancies between the data and SED templates.
Thus the broad-band SED of J0331$-$2755 does not differ 
significantly from those of typical radio-loud quasars, despite the 
double-peaked nature of this source.
The bolometric luminosity estimated based on the 
\citet{Elvis1994} template is $L_{\rm bol}\approx5.7\times10^{45}$ \lum, 
corresponding to an 
accretion rate of $\sim1$~$M_{\sun}$~yr$^{-1}$ under 
the assumption of accretion efficiency $\eta=0.1$. Note that the template SED
will `double-count' the IR emission if the IR bump consists of 
reprocessed nuclear continuum emission, which could result in an 
overestimate of the bolometric luminosity by
up to 20--30\%.

\section{Discussion}
\begin{figure}
\centerline{
\includegraphics[scale=0.5]{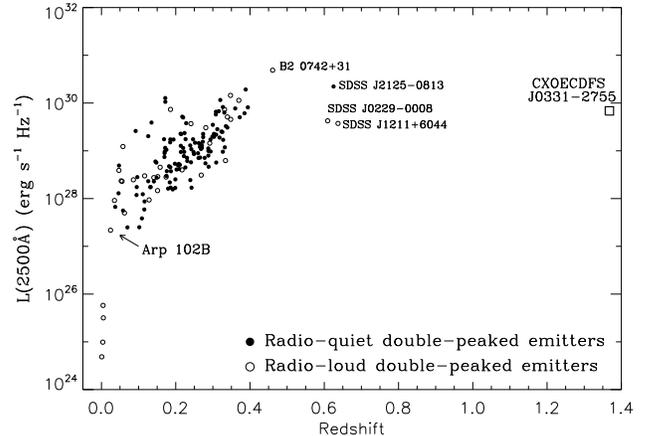}
}
\figcaption{Redshift-luminosity distribution of double-peaked emitters 
\citep[adapted from][]{Strateva2008}.
Filled dots represent radio-quiet sources, and open dots represent radio-loud 
sources. The positions of J0331$-$2755, Arp 102B, and a few high-redshift 
sources are indicated.
\label{lz}}
\end{figure}

The discovery of J0331$-$2755 doubles the redshift range of 
known double-peaked emitters, from $\sim0.6$ to $1.369$. 
The rest-frame 2500 \AA\ monochromatic luminosity of J0331$-$2755 is 
$7.2\times10^{29}$ \mlum, interpolated from the \hbox{COMBO-17} 
flux densities. It is thus among the few most optically 
luminous double-peaked emitters known.
Figure~\ref{lz} shows the position of J0331$-$2755 in the
redshift versus 2500 \AA\ monochromatic-luminosity plane. Data for the
other double-peaked emitters are from \citet{Strateva2008} 
and references therein.
The X-ray and bolometric
luminosities of J0331$-$2755 are also comparable to those for the brightest
double-peaked emitters.
The UV--to--X-ray index, 
defined as 
$\alpha_{\rm OX}=-0.3838\log[F_{\nu}(2500~{\rm \AA})/F_{\nu}(2~{\rm keV})]$, 
is $-1.27$. Compared to the $\alpha_{\rm OX}$--$L_{\rm 2500~\AA }$ relation for
typical radio-quiet AGNs in \citet{Steffen2006}, 
J0331$-$2755 has a flatter $\alpha_{\rm OX}$ 
(predicted $\alpha_{\rm OX}=-1.45$ at this $L_{\rm 2500~\AA }$). This
factor of $\sim3$ X-ray enhancement is expected
given the radio-loud nature of the source; jet-linked X-ray emission
is likely making a substantial contribution to the X-ray spectrum.

Following \citet{Eracleous1994}, we estimated the viscous 
power released in the line-emitting region $W_{\rm d}$.
Assuming an accretion efficiency 
of $\eta\approx0.1$ ($L_{\rm bol}=\eta\dot{M}c^2$), the gravitational 
power output is given by
\begin{equation}
W_{\rm d}=7.7\times L_{\rm bol}\left[\frac{1}{\zeta_1}\left(
1-\sqrt{\frac{8}{3\zeta_1}}\right)-\frac{1}{\zeta_2}\left(
1-\sqrt{\frac{8}{3\zeta_2}}\right)\right]~{\rm erg~s^{-1}},
\end{equation}
where $\zeta_1$ and $\zeta_2$ 
are the inner and outer radii of the emission region in units of
$R_{\rm G}$. For J0331$-$2755, 
$L_{\rm bol}\approx5.7\times10^{45}$~\lum, $\zeta_1\approx200$, 
$\zeta_2\approx900$, and
$W_{\rm d}\approx1.5\times10^{44}$ \lum.
The ratio of the \ion{Mg}{2} luminosity 
to the viscous power is then $L_{\rm Mg~II}/W_{\rm d}\approx0.07$. 
Assuming that 
J0331$-$2755 has the same H$\alpha$ to \ion{Mg}{2} line ratio as Arp 102B 
\citep[$\sim5.0$;][]{Halpern1996}, the H$\alpha$ luminosity is
$\sim35\%$ of the total energy available locally.
Based on computations of the emission from the accretion disks 
of cataclysmic variables in \citet{Williams1980}, we expect that 
no more than 20\% of the 
local viscous energy would be emitted as H$\alpha$. Thus the local energy
is probably insufficient to power the strong lines, and 
external illumination of 
the accretion disk appears necessary to explain the observed \ion{Mg}{2} line
luminosity even for this luminous, and likely efficiently accreting, 
double-peaked emitter.

The source of the external illumination is still uncertain.
For the prototype double-peaked emitter Arp 102B, \citet{Chen1989} proposed 
that the outer accretion disk was illuminated by a 
vertical extended structure in the inner disk, such as a geometrically thick and
optically thin X-ray-emitting flow produced by radiatively inefficient accretion
\citep[RIAF; e.g.,][]{Rees1982,Narayan1994}. This mechanism has also been
proposed to apply to other low Eddington ratio, low-luminosity 
double-peaked emitters.
However, recent studies have revealed 
that some sources are actually efficient accretors, with
Eddington ratio $L/L_{\rm Edd}\ga 0.1$, a regime where RIAFs cannot exist 
(e.g., \citealt{Lewis2006}; see also \citealt{Strateva2008}).
The SED of J0331$-$2755 shows a big blue bump (BBB) in the UV, 
as well as a typical X-ray photon index and luminosity for radio-loud quasars,
also indicating a 
relatively high-efficiency accretion flow in the central engine. 
In Figure~\ref{sed}, we show the SED of Arp~102B 
\citep[][and references therein]{Eracleous2003a,Strateva2008}.
Compared to J0331$-$2755, Arp 102B has a luminosity about two orders
of magnitude fainter and lacks a BBB.
For efficiently accreting double-peaked emitters, 
external photons may come from 
a different kind of disk-illuminating structure, for example, disk photons 
scattered by electrons in jets or slow outflows
as suggested by \citet{Cao2006}.

Another open question regarding double-peaked emitters 
is their connection to the general AGN population. 
Observationally, except for their double-peaked and generally broader 
emission lines, many double-peaked emitters resemble 
typical AGNs (e.g., through broad-band
optical--to--X-ray SEDs). 
Despite the requirement of \hbox{X-ray} illumination of the disk, their X-ray
properties (e.g., spectra and power output) do not differ 
greatly from those of single-peaked AGNs 
\citep[e.g.,][]{Strateva2003,Strateva2006}, though recent study suggested
enhanced \hbox{X-ray}
emission relative to the UV/optical emission in
a sample of the broadest double-peaked emitters \citep{Strateva2008}.
Theoretically, all luminous AGNs are expected to have accretion 
disks, and Keplerian disks 
are capable of producing double-peaked emission lines. Although the disk
parameters, such as the inclination and the ratio of the inner to outer radius, could affect the appearance of 
the line profile, they are not sufficient to explain 
why only $\sim3\%$ of AGNs are double-peaked emitters. 
\citet{Murray1997} suggested that a varying optical depth in an accretion-disk
wind could determine the presence of single or double-peaked line profiles. 
The underlying double-peaked lines become single-peaked due to radiative
transfer in a strong radiation-driven disk wind, and therefore 
double-peaked emission lines are mostly existent in low-luminosity AGNs,
where disk winds are weak with small optical depths. 
However, this model cannot 
explain the existence of the most luminous double-peaked emitters, including
J0331$-$2755.
Alternatively, the presence of single or double-peaked emission lines 
could be related to the structure
and kinematics of the broad-line region (BLR). It is commonly believed that 
the BLR is highly stratified. The low-ionization BLR could consist of two
kinematically distinct regions: an outer region of the accretion disk and 
a standard kinematically hot broad-line cloud component. If in the majority
of AGNs line emission from the latter component dominates, 
we will see single-peaked broad emission
lines, while in the remaining $\sim3\%$ we see double-peaked lines from 
the accretion disk. However, the detailed structure of the
BLR is poorly understood. 
It seems likely that the small fraction of double-peaked
emitters among AGNs is the result of a combination of these factors, i.e.,
the influence of external illumination, the position and inclination 
of the line-emitting region of the accretion disk, the presence of disk winds,
and the complex geometric nature of the BLR.
The properties of J0331$-$2755 indicate that at higher
redshift, double-peaked emitters still possess typical AGN SEDs and            
X-ray spectra, and the double-peaked line profile can also be explained 
by the Keplerian disk model.

Long-term profile variability has been found to be an ubiquitous 
property of double-peaked emitters \citep[e.g.,][]{Gezari2007}.
As the \hbox{E-CDF-S} field will continue to be surveyed in 
spectroscopic campaigns, it is likely that 
additional optical/near-IR spectroscopic data
for J0331$-$2755 can be obtained in the future. These
will help to study line-profile variability and constrain 
better the structure
and kinematics of the accretion disk, which could shed light
on some of the unresolved problems discussed above.

~\\
We acknowledge financial
support from NASA LTSA grant NAG5-13035 (BL, WNB), CXC grant SP8-9003A/B 
(BL, WNB, FEB), the Pol\'anyi Program of NKTH (GS), and the Italian Space Agency 
(contracts ASI--INAF I/023/05/0 and ASI I/088/06/0) and PRIN--MIUR grant
2006-02-5203 (CV).
We thank M.~Brusa, M.~Eracleous, B.P.~Miller, J.~Wu, 
and Y.~Xue for helpful discussions. We also thank the referee for carefully 
reviewing the manuscript and providing helpful comments.


\begin{deluxetable}{lc}

\tablewidth{0pt}

\tablecaption{SED Data for J0331$-$2755}

\tablehead{
\colhead{Band} &
\colhead{log \lnu\ (\lum)}
}

\startdata
{\it Radio} &  \\
~~~VLA 1.4 GHz & 41.27 \\
\hline
{\it Infrared} &  \\
~~~FIDEL 70 $\mu$m & $<44.88$ \\
~~~FIDEL 24 $\mu$m &44.85 \\
~~~SIMPLE 8.0 $\mu$m &44.91 \\
~~~SIMPLE 5.8 $\mu$m &44.91 \\
~~~SIMPLE 4.5 $\mu$m &44.96 \\
~~~SIMPLE 3.6 $\mu$m &44.93 \\
\hline
{\it Optical}\,\tablenotemark{a} & \\
~~~COMBO-17 I & 44.85\\
~~~COMBO-17 R & 44.95\\
~~~COMBO-17 U & 44.94\\
\hline
{\it UV} & \\
~~~{\it GALEX} 2267 \AA &45.04 \\
\hline
{\it X-ray} & \\
~~~\chandra\ 2 keV &44.35 \\
~~~\chandra\ 5 keV &44.46 \\
\enddata
\label{phodata}
\tablenotetext{a}{\,The full set of COMBO-17 SED data points are available in
\citet{Wolf2004,Wolf2008}.}
\end{deluxetable}

\end{document}